\documentstyle[12pt]{article}
 \textwidth
 16.4cm
 \oddsidemargin
 2.5cm
 \advance\oddsidemargin
 by
 -1in
 \evensidemargin
 0.0cm
 \advance\evensidemargin
 by
 -1in
 \marginparwidth
 1.9cm
 \marginparsep
 0.4cm
 \marginparpush
 0.4cm
 \topmargin
 -1.5cm
 \advance\topmargin
 by
 -0.0in
 \textheight
 23.0cm
 \makeindex

 \newcommand\noi{\noindent}
 \newcommand\beq{\begin{equation}}
 \newcommand\eeq{\end{equation}}
 \newcommand\beqn{\begin{eqnarray}}
 \newcommand\eeqn{\end{eqnarray}}
 \newcommand{\doublespace}
 {
 \newcommand{\la}{\langle}
 \newcommand{\ra}{\rangle}
 \renewcommand{\baselinestretch}
 {1.6}
 \large\normalsize}

\begin{document}

\vspace*{3cm}

\centerline{\Large \bf  $\boldmath 
J/\Psi N $ and $\boldmath \Psi'N$
total cross
sections
from}

\medskip
\centerline{\Large \bf photoproduction
data:
%
failure of vector
dominance}

\vspace{.5cm}
\begin{center}
 {\large J\"org~H\"ufner$^{1,2}$
and
Boris~Z.~Kopeliovich$^{1,2,3}$}\\
\medskip
{\sl $^1$Institut f\"ur Theoretische Physik der
Universit\"at, Philosophenweg 19, 
\\
69120
Heidelberg,
Germany}\\

{\sl $^2$ Max-Planck
Institut
f\"ur
Kernphysik,
Postfach
103980,
69029
Heidelberg,
Germany}\\


{\sl $^3$
Joint
Institute
for Nuclear Research,
Dubna,
141980
Moscow
Region,
Russia}

\end{center}

\vspace{.5cm}
\begin{abstract}

Values for the cross sections $\sigma^{\Psi N}_{tot}$
and $\sigma^{\Psi' N}_{tot}$
are of crucial importance, once one wants
to understand nuclear suppression of $J/\Psi$($\Psi'$)
in a search for the quark gluon plasma.
Quoted values are usually extracted from
$\gamma N \to J/\Psi(\Psi') N$ data via the vector
dominance model (VDM) and are very small compared with
expectations from QCD. The validity of QCD is questioned.
When the
VDM is extended to a multi-channel case, the $\gamma N$ data
lead to a value $2.8\pm 0.12\ mb <\sigma_{tot}^{\Psi N}(\sqrt{s}=
10 \ GeV) < 4.1\pm 0.15\ mb$, 
which is a factor 3 larger than the VDM prediction. 
Errors are due to experimental input, and the corridor shows the
theoretical uncertainty.
We also derive from the data
$\sigma_{tot}^{\Psi' N}/\sigma_{tot}^{J/\Psi N}\approx 4$, 
where we could not estimate the theoretical uncertainty.

\end{abstract}

\newpage

\doublespace
\noi
{\large\bf
1. Introduction}
\medskip

The success of the vector dominance model (VDM)
for
the
description of high-energy $\gamma p$
interactions,
particularly
for the production of light vector mesons
\cite{bauer}
has
led to a wide spread confidence that the VDM also accurately
describes
the
production of heavy flavoured 
vector mesons. Using data
on
elastic
photoproduction
of
charmonia,
$J/\Psi$ and $\Psi'$, the use of the VDM leads to a  
$J/\Psi$-nucleon total cross section $\sigma^{\Psi
N}_{tot}
\approx
1.3\ mb$ for $\sqrt s =
10$~GeV
and
$\sigma^{\Psi'N}_{tot}/\sigma^{\Psi
N}_{tot}\approx0.8$
(see below), which are factors 2 to 4 smaller than predicted by QCD
\cite{kz,hp}.

Reliable values for the charmonium interaction cross sections
are of
crucial importance, since $J/\Psi$ and $\Psi'$
suppressions in heavy ion collisions
are
possible
signatures for the quark-gluon plasma
(QGP)
formation.
One needs to know the cross section $\sigma^{\Psi N}_{tot}$ in
the
absence
of
a
QGP-environment in order to predict a nuclear suppression
of
$J/\Psi$,
which can 
be used as a base line in search for new physics. An
analysis
of
available data for
proton-nucleus
collisions
\cite{gh,na50} in a simple model
assuming
instantaneous
production of the $J/\Psi$ \cite{gh} (which has
a
reasonable
accuracy at medium energies $\sqrt{s} \sim 10$~GeV 
\cite{hkn}), leads to
a value of $\sigma^{\Psi
N}_{tot}
\approx 6\ mb$ \cite{ghq}, which is substantially higher 
than the VDM prediction (see also \cite{k}). 

In this paper we argue that the discrepancy between the
values
for
$\sigma^{\Psi
N}_{tot}$ from photoproduction and hadroproduction data is to be
sought
in
a
too
naive analysis of photoproduction data. 

Also a disagreement of the VDM prediction for the
$\phi
N$
cross section with the prediction of the additive
quark
model
is claimed in \cite{dl,l}. In this case
more
information
about $\sigma^{\phi N}_{tot}$ is
available
than
for
charmonium.
Facing such a problem with VDM in the strange
sector,
one
reasonably concludes \cite{l} that predictions of the VDM
for
heavier
flavours are even less trustable.

The purpose of this paper is to clarify {\it
why}
the VDM fails
for
heavy
flavours, and {\it how} it can be modified so that one
can
extract
reliable values for heavy
quarkonium interaction cross sections from
photoproduction
data.

\bigskip

{\large\bf 2. Predictions of the
VDM}

\medskip

The VDM establishes a relation between the
elastic
photoproduction
$\gamma
N\to
VN$
cross section of a vector meson $V$ and the total
cross
section
$\sigma^{VN}_{tot}$
of this meson on
a
nucleon
\cite{bauer},

\beq
\left(\sigma^{VN}_{tot}\right)_{VDM}
=
\left[\left.\frac{16\pi\alpha_{em}M_V}
{3\
\Gamma^V_{e^+e^-}(1+\rho_V^2)}\
\frac{d\sigma(\gamma N
\to
VN)}
{dt}\right|_{t=0}\right]^{1/2}\
.
\label{0}
\eeq
\noi
Here $\rho_V$ is the ratio of real to imaginary
parts of the forward $VN$ elastic scattering amplitude.
In most cases, data are available only for
the
angle
integrated cross section, which is
related
to
the forward elastic photoproduction cross
section
by
$d\sigma(\gamma N
\to
VN)/dt|_{t=0}=B_{el}^{VN}\
\sigma(\gamma N \to VN)$. 
The slope $B_{el}^{\Psi N}$
of the elastic
$J/\Psi N$
differential
cross
section
is determined mainly by the nucleon formfactor,
since
the
radius of heavy quarkonium
is
small,
$\langle r^2\rangle_{\Psi} \ll
\langle
r^2\rangle_N$.
Therefore, $B_{el}^{\Psi N}$ is only
slightly
larger
than the half of the
slope
of
elastic $pp$ scattering. Our
estimate
$B_{el}^{\Psi N}
\approx 6\ GeV^{-2}$ \cite{hkz} agrees well
with
experimental
data
\cite{chiara}.

Refs. \cite{hd97,h1,zeus} give a collection of experimental
data, which we fit with the expression,
\beq
\sigma (\gamma N\to\Psi N)
=
\sigma_0\
\left(\frac{\sqrt{s}}{10\
GeV}\right)^{2\lambda}\
,
\label{01}
\eeq
\noi
with parameters $\sigma_0 =
(10.3\pm 0.7)\
nb$,
$\lambda = 0.40 \pm 0.02$.
We can estimate $\rho_{\Psi}= (\pi/2)d[\ln(\sigma^{\Psi
N}_{tot})/d[\ln(s)] = \pi\lambda/4 \approx 0.3$.

Applying the VDM relation (\ref{0}) to the parameterization
(\ref{01})
we extract
a value\footnote{Actually, the exponent can be about $20\%$
larger than $\lambda$ due to energy dependence 
of the slope $B^{\Psi N}_{el}$}.
\beq
\left(\sigma^{\Psi
N}_{tot}\right)_{VDM}=(1.24\pm 0.13)\ mb
\times (\sqrt{s}/10\ GeV)^{\lambda}\ .
\label{01a}
\eeq

Using the VDM relation (\ref{0}) 
one can also extract
the
total
cross section
for
the
radial excitation $\Psi'(2S)$. One finds for
the
ratio
of the $\Psi'$ to $\Psi$
cross
sections

\beq
\frac{\sigma^{\Psi'
N}_{tot}}
{\sigma^{\Psi
N}_{tot}}
=
\left(\frac{M_{\Psi'}\
\Gamma^{\Psi}_{e^+e^-}}
{M_{\Psi}\
\Gamma^{\Psi'}_{e^+e^-}}\
R\right)^{1/2}
\label{02}
\eeq
\noi
with the ratio of the photoproduction
cross
sections
$R = \sigma (\gamma N \to
\Psi'
N)/
\sigma (\gamma N \to \Psi N)$. For the
experimental
value
$R=0.195\pm
0.03$
\cite{binkley}-\cite{barate}, eq.~(\ref{02}) gives 
\beq
\left(\frac{\sigma^{\Psi'
N}_{tot}}{
\sigma^{\Psi
N}_{tot}}\right)_{VDM}
=
0.8\pm 0.2\ .
\label{02a}
\eeq
\noi
This result contradicts dramatically
the
expectation
based
on
QCD, 

\beq
\frac{\sigma^{\Psi'N}_{tot}}{\sigma^{\Psi
N}_{tot}}=
\frac{<r^2>_{\Psi'}}{<r^2>_\Psi}\ .
\label{02b}
\eeq
\noi
This ratio is $7/3$ in
the harmonic oscillator model for $J/\Psi$ and
$\Psi'$, and equals 4 in a more realistic 
potential model \cite{buch}.

\bigskip

{\large\bf 3. Why does VDM fail for
heavy
flavours?}

\medskip

The basic assumption of the VDM is that via
quantum
mechanical
fluctuations
the photon converts into a vector meson $V$, which interacts
with
the
proton elastically and thereby comes on the
energy
shell. Is
the
hadronic fluctuation of the $\gamma$ always a
vector
meson?
In ref.~\cite{b-m,kz,r}
it
is
proposed that the photoproduction of $J/\Psi$ should be
considered in perturbative QCD 
as
an interaction
of
a
$c\bar c$ fluctuation (and not a $J/\Psi$ fluctuation) of the photon
with
the
target
followed by projection of the produced $c\bar c$ wave packet
on
the
$J/\Psi$ wave
function. Only
if
the wave function of the $c\bar c$ component
of
the
photon is similar to that for the $J/\Psi$, this model is close
to
the
conventional
VDM.

In order to see when and to what degree the wave functions of
$Q\bar
Q$
fluctuations
coincide with those of the vector mesons, considered as $Q\bar
Q$
bound
states,
we
study the mean squared transverse radii. The wave function
of
the
$Q\bar Q$ component of the photon has the
form
\cite{kz},

\beq
\Psi^{\gamma}_{Q\bar
Q}(r_T)
\propto
K_0(m_Qr_T)\
,
\label{1}
\eeq
\noi
where $K_0(x)$ is the modified Bessel function and
$r_T$
is
the transverse $Q\bar
Q$
separation.
Its mean transverse size squared is calculated
from
(\ref{1})

\beq
\langle
r_T^2\rangle_{\gamma}=
{2\over
3m_Q^2}\ .
\label{2}
\eeq
This value has to be compared with the mean
transverse
distance
between the $Q$ and $\bar Q$ in the vector meson, considered as
the
lowest
state
in
an harmonic oscillator potential with
frequency
$\omega$:

\beq
\langle
r_T^2\rangle_V
=
\frac{2}{m_Q\
\omega}\
.
\label{3}
\eeq
\noi
Empirically one finds that $\omega
=
(M_{V'(2S)}-M_{V(1S)})/2$
has a value of roughly 300~MeV independent of
the flavour. 

The two sizes eqs.~(\ref{2}) and (\ref{3}) 
depend differently on $m_Q$, they cannot coincide
for all flavours, in particular, the
discrepancy increases the heavier the quark mass is.

We introduce the factor $N_{\Psi}$ and $N_{\Psi'}$
for the transition from the VDM prediction to the real value
\beq
\sigma^{\Psi N}_{tot}=
{1\over N_{\Psi}}
\left(\sigma^{\Psi N}_{tot}\right)_{VDM}
\label{3a}
\eeq
\noi
(and similarly for $\Psi'$) and attempt to calculate 
these factors by two different methods in the next sections.

\bigskip

{\large\bf 4. Hadronic representation: a multi-channel approach}

\medskip

The above observation that the $c\bar c$
fluctuation
is
quite different from
the
$J/\Psi$
can be phrased formally in writing an expansion of the $c\bar
c$
fluctuation
in
terms of a complete set of
charmonium
states
\beq
|\gamma\rangle_{c\bar c}
=\alpha_1|J/\Psi\rangle + 
\alpha_2|\Psi'\rangle+...\ .
\label{4a}
\eeq
\noi
With this expansion of the $c\bar
c$
fluctuation,
photoproduction of $\Psi'$ and $J/\Psi$ gains contributions
from
different
mechanisms as is shown schematically
in
Fig.~2 for a two channel example:
i) direct production as in the VDM $\gamma\to\Psi,\Psi N\to\Psi
N$
and
ii) indirect production, $\gamma\to\Psi',\Psi'N\to\Psi N$, which
contains
the
off-diagonal diffractive interaction $\Psi N\rightleftharpoons
\Psi'
N$.
The two amplitudes which lead to the final state must be
added
before
squaring.
The non-diagonal process $\Psi\to\Psi'$ is the correction to
the
VDM
expression
which
we
propose.

\begin{figure}[tbh]
\includegraphics{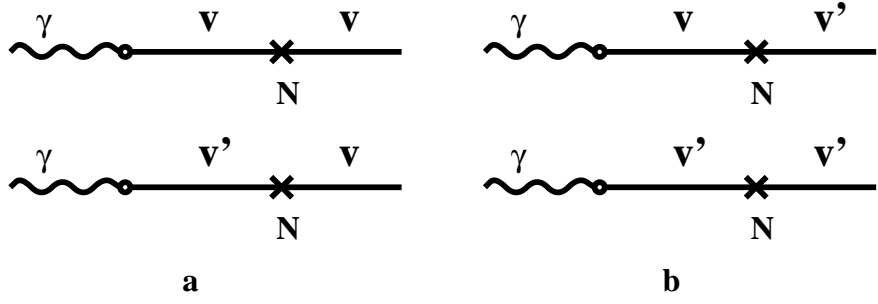}
\begin{center}
\vspace{4cm}
\parbox{13,3cm}
{\caption[Delta]
{\sl Diagrams for the elastic
photoproduction
of the vector mesons, an example of two channels.
Photoproduction of the ground state $V$ ({\bf a}),
and of the radial excitation $V'$ ({\bf b}).}
\label{vv}}
\end{center}
\end{figure}

The virtual photoproduction amplitude 
in the naive VDM reads,

\beq
f_{VDM}(\gamma^* N\to \Psi
N)
=
\frac{C^{\gamma\Psi}}
{M_{\Psi}^2+Q^2}\ f(\Psi N\to
\Psi
N)
\
,
\label{11}
\eeq
\noi
where the photon of virtuality $Q^2$ converts into $\Psi$ with
a
probability
amplitude $C^{\gamma\Psi}$, the $J/\Psi$ propagates
with
a
usual propagator and $f(\Psi N \to \Psi N)$
denotes
the
elastic scattering amplitude. The
extension
of
eq.~(\ref{11}) to the multi-channel case
is
straightforward:

\beq
f(\gamma^* N\to \Psi
N)
=
N_{\Psi}(Q^2)\
f_{VDM}(\gamma^* N\to
\Psi
N)
\
,
\label{12}
\eeq
\noi
where we have defined the correction
factor
$N_{\Psi}(Q^2)$

\beq
N_{\Psi}(Q^2) = 1
+
\sum\limits_{i\geq 1}
\frac{C^{\gamma\Psi_i}\ f(\Psi_i N\to
\Psi
N)}
{C^{\gamma\Psi}\ f(\Psi N\to
\Psi
N)}\
\frac{M_{\Psi}^2
+
Q^2}
{M_{\Psi_i}^2
+
Q^2}\ .
\label{13}
\eeq

It follows from perturbative QCD (and is shown in the next
section) that $N_{\Psi}\propto 1/Q^2$ for $Q^2 \to \infty$.
This property of color transparency 
implies for the expression (\ref{13}) a sum
rule,
\beq
\sum\limits_{i\geq 1}
\frac{C^{\gamma\Psi_i}\ f(\Psi_i N\to
\Psi
N)}
{C^{\gamma\Psi}\ f(\Psi N\to
\Psi
N)}\
=
-
1\ .
\label{14}
\eeq
\noi
For heavy quarkonia the mass splitting is much smaller 
than the mass, therefore, we expect a strong cancellation in 
(\ref{13}) even at low values of $Q^2$. Indeed, the zero order term in
an expansion of $N_{\Psi}(0)$ in the small parameter
$\Delta M_{\Psi}/M_{\Psi}$ vanishes according to (\ref{14}).
The first nonzero term in the expansion is of the order of
\beq
N_{\psi}(0) \sim
\frac{\Delta
M_{\Psi}}
{M_{\Psi}}\approx 0.2\ .
\label{15}
\eeq

The observed strong cancellation between
different
channels
in (\ref{12}) is a general property of
heavy
quarkonia,
and we do not expect the factor $N_{\Psi}(Q^2)$ to
depend
much on the set of orthogonal
states
chosen for the hadronic
basis.
In one case the
function
$N_{\Psi}(Q^2)$ can be calculated exactly.
If
the
operator
for the
elastic
scattering amplitude of a $c\bar c$ pair on a
nucleon
is proportional to the transverse quark
separation
squared
$r_T^2$, and harmonic oscillator wave functions
are
chosen for the hadronic
basis, one has $f(\Psi_i N\to\Psi N)\propto\langle\Psi_i|
r_T^2|\Psi_0\rangle=0$
for $i\geq 2$. Then, according to the sum rule (\ref{14}),
one finds the surprisingly simple result,
\beq
N_{\Psi}(Q^2)
=
\frac{M_{\Psi'}^2
-
M_{\Psi}^2
}
{M_{\Psi'}^2 +
Q^2}\
,
\label{15a}
\eeq
\noi
which is independent of any matrix element.
With $N_{\Psi}(0)= 0.30$, one arrives at a 
factor of $3.3$ correction to the naive VDM,
\beq
\sigma^{\Psi N}_{tot} = 
(4.3 \pm 0.5)\ mb\ 
\times\left(\frac{\sqrt{s}}
{10\ GeV}\right)^{0.4}\ .
\label{16a}
\eeq

How good in the correction factor (\ref{15a})?
We will give an independent calculation of $N_{\Psi}(Q^2)$ 
in the next sections. In
fig.~2 we compare
our predictions with the factor
$N_{\Psi}(Q^2)$ Eq.~({\ref{15a}) and with $N_{\Psi}(Q^2)=1$,
which is the conventional VDM,
 and available data from the EMC
experiment
\cite{aubert}. 
\begin{figure}[tbh]
\includegraphics{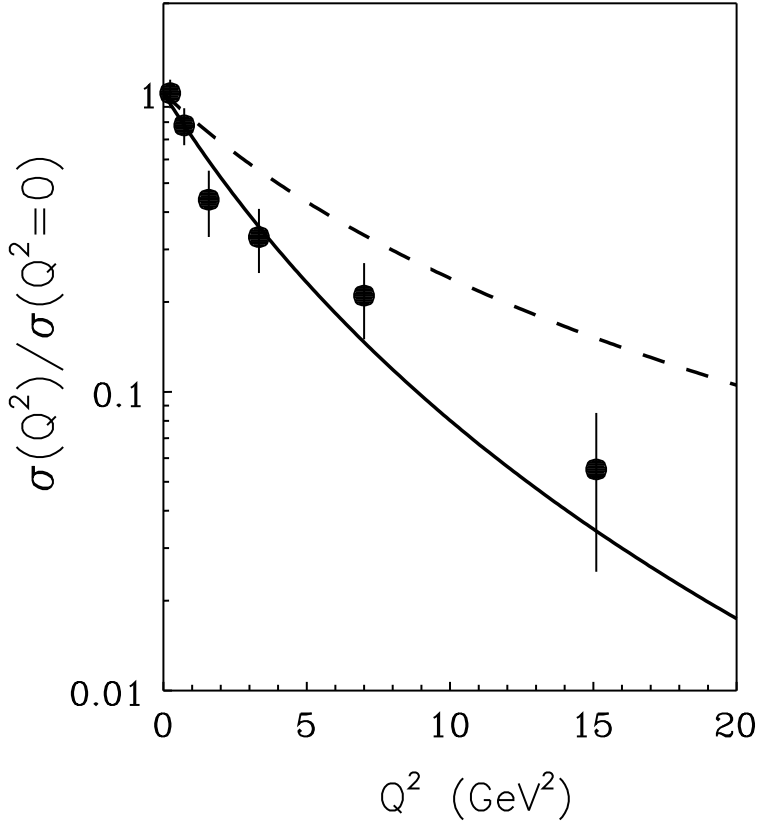}
\begin{center}
\vspace{10.5cm}
\parbox{13cm}
{\caption[Delta]
{\sl Data on the $Q^2$-dependence of the exclusive 
muoproduction of $J/\Psi$
 \cite{aubert}. The dashed curve is the
expectation of the VDM,
the solid curve includes the correction 
factor (\ref{15a}).}
\label{fig2}}
\end{center}
\end{figure}
The comparison obviously supports
the
form (\ref{15a}), although the error bars are rather
large.

For the $\Psi'$ production we have to introduce
at least three
intermediate states $\Psi$, $\Psi'$ and $\Psi''$,
which
exhaust the sum in the hadronic oscillator basis and
have

\beqn
&&f(\gamma^*p\to\Psi'p)=
f_{VDM}(\gamma^*p\to\Psi'p)\nonumber\\
&\times&\left[ 1 +
\sum\limits_{i=0,2}
\frac{C^{\gamma\Psi_i}\
f(\Psi_iN\to\Psi'N)}
{C^{\gamma\Psi'}\
f(\Psi'N\to\Psi'N)}\
\frac{M_{\Psi'}^2+Q^2}{M_{\Psi_i}^2+Q^2}
\right]\
,
\label{17}
\eeqn
where the squared bracket defines
$N_{\Psi'}(Q^2)$. In an 
oscillatory basis $f(\Psi' N\to\Psi' N)/f(\Psi
N\to\Psi
N)=7/3$. Together with the color transparency sum rule
(\ref{14}) all the
parameters are determined

\beq
N_{\Psi'}(Q^2)
=
1 -
\frac{2}{7}\
\frac{M_{\Psi'}^2+Q^2}
{M_{\Psi}^2+Q^2}
-
\frac{5}{7}\
\frac{M_{\Psi'}^2+Q^2}
{M_{\Psi''}^2+Q^2}\
,
\label{18}
\eeq
which expression leads to
$N_{\Psi'}(0)=0.064$.
Then the ratio
of the corrected values for the 
$\Psi'N$ to $\Psi N$ total cross sections is 
\beq
\frac{\sigma^{\Psi'N}_{tot}}
{\sigma^{\Psi N}_{tot}}
=
\frac{N_{\Psi}(0)}
{N_{\Psi'}(0)}\
\left(
\frac{\sigma^{\Psi'N}_{tot}}
{\sigma^{\Psi
N}_{tot}}
\right)_{VDM} = 3.75\
.
\label{19}
\eeq
\noi
This ratio corresponds nearly exactly
to the QCD prediction, namely, 
the ratio of radii squared, which calculated for a
realistic
potential
\cite{buch} gives a value 4.

\bigskip

{\large\bf 5. Quark representation}

\medskip

In order to estimate the theoretical uncertainty of the correction
factor $N_{\Psi}(0)$ and $N_{\Psi'}(0)$ evaluated using the harmonic oscillator basis, we calculate these factors by yet another approach.

The multi-channel equations (\ref{12}) -
(\ref{13}) can be represented in the quark
basis as,
\beq
f(\gamma^*N \to \Psi N) = \langle \Psi|
\sigma(r_T)|\gamma^*_{c\bar c}\rangle
\label{19b}
\eeq
\noi
with the wave function for the photon $|\gamma^*_{c\bar c}\rangle$
in the quark basis. In the same representation, the expression
for the VDM reads
\beq
f(\gamma^*N \to \Psi N) = \langle \Psi|
\sigma(r_T)|\Psi\rangle\langle\Psi|
\gamma^*_{c\bar c}\rangle\ .
\label{19ba}
\eeq

Then the correction factor takes the form
\beq
N_{\Psi}(Q^2)= \frac{
\langle \Psi|
\sigma(r_T)|\gamma^*_{c\bar c}\rangle}
{\langle \Psi|
\sigma(r_T)\Psi\rangle\langle\Psi|
\gamma^*_{c\bar c}\rangle}\ ,
\label{19bb}
\eeq
\noi
where $\Psi$ may stand for the wave functions of the $J/\Psi$ or
the $\Psi'$. The perturbative 
 wave function of the $c\bar c$ fluctuation
of the virtual photon reads in non-relativistic
approximation, $|\gamma^*_{c\bar c}\rangle \propto
K_0(\epsilon r_T)$ \cite{nz91,kz}, where $\epsilon^2=m_c^2+Q^2/4$.
This perturbative QCD expression for $|\gamma^*_{c\bar c}\rangle$
neglects the interaction between $c$ and $\bar c$ and therefore is 
questionable for low $Q^2$. In this respect the hadronic 
representation is preferable, since effectively it takes into account
 all the non-perturbative effects.

The mean squared $c\bar c$ separation in the photon is 
calculated to $\langle r_T^2\rangle_{\gamma^*} = 2/3(m_c^2+Q^2/4)$
and goes to zero for $Q^2\to\infty$.
In this limit and using $\sigma(r_T)=ar_T^2$, the expression for 
$N_{\Psi}(Q^2)$ becomes very simple, because 
$\Psi(r_T)$ can be replaced by $\Psi(0)$ in the integrals involving
$K_0(r_T)$.
Then,
\beq
N_{\Psi} \to
\frac{\langle r_T^2\rangle_{\gamma^*}}
{\langle r_T^2\rangle_{\Psi}}
\hspace{1.5cm}
(Q^2\to\infty)\ .
\label{19bc}
\eeq
\noi
This result is exact. We note, that $\langle r^2_T\rangle_{\gamma^*}$
goes to zero with the $Q^2 \to \infty$ and this leads to the sum rule
(\ref{14}).
If one uses this expression
also at $Q^2=0$ together with $\langle r_T^2\rangle_{\Psi}=
2/\omega m_c$ eq.~(\ref{3}) one finds $N_{\Psi}(0) = 2\omega/m_c$,
which agrees with (\ref{15a}) to lowest order in $\omega/2m_c$.
In the same limit 
\beq
\frac{N_{\Psi'}(Q^2)}
{N_{\Psi}(Q^2)} = 
\frac{\langle r_T^2\rangle_{\Psi'}}
{\langle r_T^2\rangle_{\Psi}}\ .
\label{19bd}
\eeq

The expression of $N_{\Psi}(Q^2)$ in (\ref{19bc})
is based on
the approximation $\langle r_T^2\rangle_{\gamma} 
\ll \langle r^2\rangle_{\Psi}$,
which is not well satisfied for charmonium. 
In order to test the accuracy, we give up this approximation, 
and use
another trial wave functions $|\Psi\rangle$, among them
the harmonic oscillator ones.

Be specific, we use a mixture of the harmonic
oscillator (HO) and Coulomb (C) wave functions,
\beq
\langle\vec r|\Psi\rangle =
\alpha \langle\vec r|\Psi_{HO}\rangle
+ \beta 
\langle\vec r|\Psi_C\rangle
\label{19c}
\eeq
\noi
for $1S$ and $2S$ states.
The Coulomb wave function is calculated with 
a running QCD coupling as is explained 
in \cite{jan}. We vary $\alpha$ ($\beta=
\sqrt{1-\alpha^2}$) and adjust the parameters
of $|\Psi_{HO}\rangle$ and $|\Psi_{C}\rangle$
so that one has always the same 
$\langle r_T^2\rangle_{\Psi}$ as taken from \cite{buch}.

Then we obtain the following results: 

\noi{\it i)} for the $J/\Psi$
\beq
2.3 \leq N^{-1}_{\Psi} \leq 3.3\ ,
\label{19ca}
\eeq
\noi
where the upper limit corresponds to (\ref{15a})
and the lower one is calculated from eq.~(19bb})
with $\alpha=0$ in (\ref{19c}). The case of pure 
Coulomb wave function ($\alpha=0$) lies in the interval.
Thus,
\beq
2.8\pm 0.12 \leq \sigma_{tot}^{\Psi N}(\sqrt{s}=10\ GeV)
\leq (4.1 \pm 0.15)\ mb
\label{19cb}
\eeq
\noi
with the energy dependence
\beq
\sigma_{tot}^{\Psi N}(\sqrt{s})=
\sigma_{tot}^{\Psi N}(10\ GeV)
\left(\frac{\sqrt{s}}{10\ GeV}
\right)^{0.4}
\label{19cd}
\eeq

\noi{\it ii)} 
The results for $N_{\Psi'}(0)$ are very sensitive to the wave function 
(\ref{19c}) (because of the node), so that we prefer
to give the result in the form (\ref{19}) without a corridor of theoretical
uncertainty.

Furthermore, we test the sensitivity of the results 
to the form of the transition operator $\sigma(r_T)$.
We choose the form
$\sigma(r_T)=a[1-\exp(-br_T^2)]/b$ \cite{jan}, which gives
the linear $r_T^2$ dependence at small $r_T$.
We found a very small deviation of the order of few percent.

\bigskip

{\large\bf 6. Conclusion
and
discussions}

\medskip

The conventional expressions of the VDM applied to
the
photoproduction
data
of
charmonium production lead to inconsistencies both with
experimental
data
($\sigma
^{\Psi N}_{tot}$ values extracted from $hA$ data) as well as
predictions
from
QCD
($\sigma^{hN}_{tot}\propto\langle r^2\rangle_h$). A generalization of the VDM
to
a
multi-channel problem leads to a value $\sigma_{tot}^{\Psi N}$
about three times as large as the one predicted by VDM,
and the
discrepancies relating to $\sigma_{tot}^{\Psi N}$ are
largely
removed.

Few more comments are
in
order.

\begin{itemize}
\item Theoretical accuracy of the correction factor $N_{\Psi}(Q^2)$
is estimated in the quark representation by varying the shape of
the $J/\Psi$ wave function and the form of the dipole cross section.
We conclude that the uncertainty does not exceed $20\%$.

\item Another source of information about $\sigma^{\Psi N}_{tot}$ is
production of $J/\Psi$ off nuclei, which give a slightly higher
value than we predict. This is because the simplified analysis \cite{na50}
does not take into account extension of the charmonium
length of path in nuclear matter, due to the coherence length \cite{hkn},
which is about $2\ fm$ for the kinematics of \cite{na50}.
Therefore a reduction of about $20\%$ should be applied to 
$\sigma^{\Psi N}_{abs}$. Besides, dynamics of charmonium hadroproduction
\cite{k} is far more complicate than for photoproduction.

In order to be free of the effects of the coherence and formation
lengths one should produce a low energy charmonium off nuclei.
A unique opportunity is annihilation $\bar p p \to \Psi$ on a bound
protons.

\item
Using the same formulas for
photoproduction
of
bottomia we arrive at
the
result,
$\sigma^{\Upsilon
N}_{tot}
=
8\
\left(\sigma^{\Upsilon
N}_{tot}\right)_{VDM}$.
The cross section
$\left(\sigma^{\Upsilon
N}_{tot}\right)_{VDM}$
as well as the
ratio
$r_{\Upsilon}$
cannot be
predicted,
since
no data on bottomium photoproduction is
available
yet.

\item
We expect a weak rising energy-dependence of the correction
factor $N_{\Psi}(Q^2)$. Indeed, it is known from HERA data on
the proton structure function $F_2(x,Q^2)$ that the smaller is the
size of a photon fluctuation, the steeper is the energy dependence.
This is why the $J/\Psi$ photoproduction cross section growth 
so steeply compared to what is known from soft hadronic interactions.
According to the QCD evolution equations this effect is due to
fast grows of the gluon cloud of the $c\bar c$ fluctuation 
of the photon, which steeper than that of $J/\Psi$.

\item
One may wonder why VDM works so well for
$\rho$
mesons.
Indeed, the same expression
(\ref{15a})
leads to deviation of $40\%$ from
the
VDM
prediction. This estimate contains an additional uncertainty
since the dipole approximation
$\sigma(r_T) \propto r_T^2$ is justified only at small $r_T$,
while large distances are important for $\rho$ mesons.
A harmonic oscillator is also too rough an 
approximation in this case.

\item
One should be cautious applying VDM to
the
photoproduction
of $J/\Psi$ on nucleons at low energy. If the lifetime
of
the
$J/\Psi$ fluctuation in the photon is much shorter than
the
nucleon
radius, $t_c \approx M_{\Psi}^2/2E_{\gamma}
\ll
R_N$,
the photoproduction cross section should
be
much
smaller than VDM predicts, since the fluctuation can
interact
only
during its short lifetime. This effect is taken
into
account
by the nucleon form factor (compare
with \cite{kp}).

\end{itemize}

\bigskip

\noi {\bf Acknowledgement}: We are thankful to
M.G.~Ryskin and B.G.~Zakharov for useful 
discussions and suggestions.
This work has been supported
by
a
grant
from
the
Gesellschaft f\"ur Schwerionenforschung Darmstadt (GSI), grant
no. HD
H\"U
T,
and
the Bundesministerium f\"ur Bildung, Wissenschaft,
Forschung
und
Technologie
(BMBF),
grant no. 06
HD
856.

\setlength{\baselineskip}
{20pt}

\end{document}